\documentclass[a4paper,11pt]{article}
\usepackage{jcappub} 

\usepackage{lipsum}
\usepackage{tabularx}
\usepackage{rotating}
\usepackage{comment}
\usepackage{xcolor}
\usepackage{soul}

\definecolor{VividBurgundy}{RGB}{159,29,53}
\newcommand*{\rom}[1]{\MakeUppercase{\romannumeral #1}}

\title{\boldmath Investigation of the Radial Profile of Galactic Magnetic Fields using Rotation Measure of Background Quasars}

\author[a]{Shivam Burman,}
\author[a]{Paras Sharma,}
\author[b,c]{Sunil Malik}
\author[a]{and Suprit Singh}

\affiliation[a]{Indian Institute of Technology Delhi, \\Hauz Khas, New Delhi, Delhi 110016, India}
\affiliation[b]{Institute fur Physik und Astronomie Universitat Potsdam, \\Golm Haus 28, D-14476 Potsdam, Germany}
\affiliation[c]{Deutsches Elektronen-Synchrotron DESY, \\Platanenallee 6, 15738 Zeuthen, Germany}

\emailAdd{Shivam.Burman@physics.iitd.ac.in}
\emailAdd{sharmaparasiitd@gmail.com}
\emailAdd{sunil.malik@uni-potsdam.de}
\emailAdd{suprit@iitd.ac.in}

\abstract{Probing magnetic fields in high-redshift galactic systems is crucial to investigate galactic dynamics and evolution. Utilizing the rotation measure of the background quasars, we have developed a radial profile of the magnetic field in a typical high-$z$ galaxy. We have compiled a catalog of 59 confirmed quasar sightlines, having one intervening Mg \rom{2} absorber in the redshift range $0.372\leq z_{\text{abs}} \leq 0.8$. The presence of the foreground galaxy is ensured by comparing the photometric and spectroscopic redshifts within $3 \sigma_{z-\text{photo}}$ and visual checks. These quasar line-of-sights (LoS) pass through various impact parameters (D) up to $160$ kpc, covering the circumgalactic medium of a typical Milky-Way type galaxy. Utilizing the residual rotation measure (RRM) of these sightlines, we estimated the excess in RRM dispersion, $\sigma_{\text{ex}}^{\text{RRM}}$. We found $\sigma_{\text{ex}}^{\text{RRM}}$ decreases with increasing D. We translated $\sigma_{\text{ex}}^{\text{RRM}}$ to average LoS magnetic field strength, $\langle B_{\|}\rangle$ by considering a typical electron column density. Consequently, the decreasing trend is sustained in the magnetic field. In particular for sightlines with $\text{D} \leq 50$ kpc and $\text{D} > 50$ kpc, $\langle B_{\|}\rangle$ is found to be $2.39 \pm 0.7 \ \mu$G and $1.67 \pm 0.38 \ \mu$G, respectively. This suggests a clear indication of varying magnetic field from the disk to the circumgalactic medium. This work provides a methodology that, when applied to ongoing and future radio polarisation surveys such as LOFAR and SKA, promises to significantly enhance our understanding of magnetic field mapping in galactic systems. 
}

\begin{document}
\maketitle{}
\flushbottom

\section{Introduction} \label{sec:intro}

Galactic magnetic fields play a pivotal role in a variety of astrophysical phenomena, such as governing star formation, influencing the dynamics and stability of the interstellar medium, and even in instigating galactic winds \citep[see refs.][]{ruzmaikin1988magnetic, shu1992physics, beck2005magnetic, brandenburg2005magnetic, krumholz2011stars, crutcher2012magnetic, 2013R.Beck, Ulrich}. An important role in the galactic evolution and dynamics is also played by the inflow and outflow of material from a galaxy to its circumgalactic medium (CGM) \citep[see the review by][]{Tumlinson_2017}. The CGM remains one of the least constrained parts of the galactic systems, especially with regard to the magnetic fields. At high redshifts, the galactic magnetic fields gain a cosmological perspective with questions such as the generation \citep[see refs.][]{rees1987formation, Kulsrud_1997, Seta_2020MNRAS} and sustenance mechanisms of magnetic fields in the galactic systems over several epochs during their evolution \citep{article_origin, subramanian2008magnetizing, Rieder_10.1093/mnras/stv2985}. It is, therefore, vital to probe the strength, structure, and evolution of magnetic fields in galaxies and their CGM to gain insights into galactic dynamics and evolution. 

There are numerous investigations on the structure and strength of magnetic fields in the Milky Way and the nearby galaxies such as M31 \citep[see refs.][]{Jansson_2012, han_2012, M31_fields, Beck_2015, Beck_IC342, sergey2018structure, seta_federrath_2021, Dickey_2022, surgent2023structure}. The average magnetic field strength over the galactic scales ranges from 0.5 to 2.5 $\mu \text{G}$ and follows a decreasing trend towards the galaxy outskirts (up to $\sim 25 \text{kpc}$). Studies on the CGM of galaxies have shown that the CGM hosts magnetic fields up to significantly large distances up to 1000 kpc from the galactic centers \citep[see refs.][]{Irwin_2016MNRAS.458.4210H, Mora_halo_2019, Voort_2021, thomas2022cosmic, Heesen_CGM_magfield,Bockmann_2023}. In addition, \cite{Heesen_CGM_magfield} have studied the CGM magnetic fields in the nearby galaxies (lying within 3.2-109 Mpc distance from the Milky Way), having an impact parameter (transverse distance from galaxy center to the line of sight) distribution up to 1000 kpc, using LOFAR observations at 144 MHz. The authors estimated the average strength of the line-of-sight (LoS) component of the magnetic field of 0.5$\mu \text{G}$. 

For the galaxies at high redshifts, particularly the Rotation Measures (RM) of polarized radio quasars have been employed extensively to investigate the magnetic fields whenever these systems occur along the quasar sightlines \citep[][]{kronberg_absorption_1982, kronberg_global_2008}. Being an integrated LoS effect, the observed quasar $\text{RM}_\text{obs}$ includes the intrinsic contribution of (i) the source and its host itself $(\text{RM}_{\text{qso}})$, (ii) the intergalactic medium (IGM) $(\text{RM}_{\text{IGM}})$,  (iii) the intervening galactic systems $(\text{RM}_{\text{intrv}})$, and (iv) the Milky Way $(\text{GRM})$ all added up. Note that RM can be negative or positive as it is given by the line integral of the magnetic field along the LoS. The occurrence of intervening systems in the LoS is detected through the absorption lines in the quasar spectra. In particular, the Mg \rom{2} absorption doublets 2796, 2803 \AA~have shown a high degree of correlation with the normal galaxies occurring at the same redshifts as these absorbers \citep[see refs.][]{Churchill_2000, Nestor_2005, Mshar_2007, Narayanan_2008, Tinker_Chen_2010}. These Mg \rom{2} absorbers can be traced up to the extent greater than $100$ Kpc from the center of galaxy \citep[see refs.][]{Kacprzak_2008, Chen_2010_MgII_extent, Bordoloi_2011, nielsen_magiicat_2013}. An excess RM has been observed for all the quasar sightlines with Mg \rom{2} absorbers in the spectra compared to those showing no absorbers. In order to obtain the RM contribution from the intervening systems, all other contributions need to be removed carefully, as there are also directional dependencies, particularly in the GRM. The estimates of the magnetic fields in the intervening systems are thus obtained from the Residual Rotation Measure (RRM = RM - GRM) statistics where the GRM contribution has been removed \citep[see refs.][]{Bergeron_Boisse_1191_MgII, kronberg_global_2008, Bernet_2010, Bernet2012, hammond2013new, joshi_dependence_2013, Farnes_2014,kim_2016}. Using these techniques, for example, \cite{malik_role_2020} gave an estimated average LoS magnetic field to be around $1.3 \pm 0.3 \mu \mathrm{G}$. Recently an article by \cite{Bockmann_2023} appeared where the authors have examined the magnetic fields properties in the CGM of 125 star-forming galaxies in the COSMOS\footnote{\href{https://cosmos.astro.caltech.edu/}{https://cosmos.astro.caltech.edu/}} (81 galaxies having spectroscopic redshift in the range 0.25-0.49) and XMM-LSS\footnote{\href{https://cesam.lam.fr/xmm-lss/}{https://cesam.lam.fr/xmm-lss/}} (44 galaxies having spectroscopic redshift in the range 0.06-1.09) fields, using the polarization data from the MeerKAT International GHz Tiered Extragalactic Exploration polarization (MIGHTEE-POL) survey, extensively provided by \cite{taylor2023mightee}. The authors obtained a redshift corrected excess $|\mathrm{RM}|$ of $5.6 \pm 2.3$ $\text{ rad } \text{m}^{-2}$ for the galaxies having impact parameter less than 133 kpc from the background radio source, with respect to the galaxies having impact parameter between the range 133-400 kpc.

While models and simulations of the structural aspects of magnetic fields in galaxies and their halos exist \citep[see refs.][]{Pakmor_CGM_2020MNRAS, thomas2022cosmic, jung2023sampling} in the literature, these structures remain largely unexplored from an observational standpoint. In one study \cite{bernet_extent_2013} analyzed 28 sources with RM values at 6 cm and concluded that the sightlines with higher values of RM have small impact parameters, following a decreasing profile up to $\sim$ 50 kpc with $\mathcal{O}(B_{||})\sim$ 10 $\mu$G. Also, \cite{Lan_and_Prochaska_2020} have reported the analysis of around 1000 high-$z$ galaxies with impact parameters up to 200 kpc and found that the RM is uncorrelated with the number of intervening galaxies along the LoS, and constrained the CGM magnetic field to be less than $2 \mu$G. The question lies open as to what is the structure and profile of the magnetic fields in these high-$z$
galaxies?

We consider this question statistically by first identifying an ensemble of quasar-galaxy pairs through Mg \rom{2} absorbers and a visual check in the Sloan Digital Sky Survey (SDSS) catalogs. Using the angular separation between the quasar LoS and the center of the intervening galaxies, we can compute the impact parameters of the sightlines. We obtain the excess contribution in RRM ($\sigma_{\text{ex}}^{\text{RRM}}$) due to the intervening galaxy by comparing the dispersion of RRM distributions of the quasar sightlines with and without absorbers. The column average strength of the LoS component of the magnetic field can then be estimated from this excess RRM. We can then answer the question posed above by seeking the correlation of the obtained magnetic fields with the impact parameters of the sources by assuming all the intervening galaxies to be similar, marginalizing their properties as a preliminary stride in the direction of profiling the magnetic fields in these galactic systems.

This paper is organized as follows. We begin with characterizing the dataset and building our sample in Section~\ref{sec:data_and_sample}. The details of our analysis, including the estimation of the impact parameter of the quasar-galaxy pairs, the radial profile of rotation measure, and the estimated magnetic field, are given in Section \ref{sec:analysis_results}. A detailed discussion of our results, as well as a comparison with previous studies, is given in Section \ref{sec:discussion}. We conclude the paper by summarizing our findings in Section \ref{sec:conclusion}.

\begin{table}
  \centering
  \resizebox{0.75\textwidth}{!}{%
  \begin{tabular}{c|l}
    \hline
    \textbf{Symbol} & \textbf{Description} \\
    \hline
    \text{RM} & Rotation measure \\
    \text{GRM} & Galactic Rotation measure \\
    \text{IGM} & Intergalactic medium \\
    $\text{RM}_\text{qso}$ & Rotation measure of host quasar \\
    \text{CGM} & Circumgalactic Medium \\
    \text{RRM} & Residual Rotation measure \\
    $B_{||}$ & Magnetic field component parallel to the line of sight \\
    $z_{\text{qso}}$ & Redshift of the background quasar \\
    $z_{\text{abs}}$ & Spectroscopic redshift of the MgII absorber \\
    $z_{\text{gal(photo)}}$ & Photometric redshift of intervening galaxy \\
    $b$ & Galactic latitude \\
    $\sigma_{z-\text{photo}}$ & Uncertainty in photometric redshift \\
    $\theta$ & Angular separation \\
    $\Delta\delta$ & Declination separation among the background quasar and absorber galaxy \\ 
    $\Delta\alpha$ & Right ascension separation among the background quasar and absorber galaxy \\
    $\delta_{\text{qso}}$ & Declination of the background quasar \\
    D & Impact parameter (in kpc) \\
    $d_A$ & Angular diameter distance \\
    $H_0$ & Hubble constant \\
    $\Omega_m$ & Matter density \\
    $\Omega_\Lambda$ & Cosmological constant \\ 
    $\sigma_{\text{abs}}^{\text{RRM}}$ & RRM dispersion of absorbers in each D bin \\
    $\text{RRM}_{\text{non-abs}}$ & RRM of the non-absorber sightlines \\
    
    $\sigma_{\text{non-abs}}^{\text{RRM}}$ & RRM dispersion of non-absorber sightlines \\
    
    $\delta\sigma_{\text{abs}}^{\text{RRM}}$ & Uncertainty in the RRM dispersion of the absorbers \\ 
    
    $\sigma_{\text{ex}}^{\text{RRM}}$ & Excess RRM due to intervening absorbers for each D bin \\
    
    $\delta\sigma_{\text{ex}}^{\text{RRM}}$ & Uncertainty in the $\sigma_{\text{ex}}^{\text{RRM}}$ \\
    $N_e$ & Column density of electrons \\
    ${\sigma_{\text{abs (sim)}}^{\text{RRM}}}$ & Excess RRM estimated using simulated data \\
    
    \hline
  \end{tabular}
  }
  \label{table:symbols}
  \caption{Table of Physical quantities used in this article.}
\end{table}

\section{Data and Sample Creation}\label{sec:data_and_sample}

In our study, we have employed the catalog of 1132 well-studied quasars compiled by \cite{malik_role_2020} derived from a positional cross-match of Sloan Digital Sky Survey Data Release (SDSS DR)-7, 9, 12, 14 quasar catalogs \citep{SDSS_DR7, SDSS_DR9, SDSS_DR12, SDSS_DR14} with the NRAO VLA Sky Survey (NVSS) RM-21cm catalog \citep{taylor_rotation_2009}. In this compilation, intervening galaxies along the LoS are reported through the Mg \rom{2} $\lambda 2796,2803$ \AA~doublet absorption lines in the quasar spectra. This criterion, combined with the wavelength coverage of SDSS, 3800-9000 \AA, selects only the absorbers lying within the redshift range of $0.372 \leq z_{\mathrm{abs}} \leq 2.2$. A positional cross-match of the catalogs has been performed with a vicinity radius of $7''$ due to the different spatial resolutions of the optical and radio telescopes. This threshold radius is identified based on localization uncertainties of telescopes involved in these catalogs and has also been investigated in \cite{Kimball_2008} and \cite{Singh_2018}. The compilation contains all the necessary observed and derived quantities, including the RRM where the contribution of the RM from the Milky Way has been removed. To segregate the sightlines based on the presence and the absence of the intervening Mg \rom{2} absorbers, the authors have also used visual identification by investigating the individual spectrum of the quasars. The catalog comprises of 352 sightlines with single and multiple intervening Mg \rom{2} absorbers and 780 sightlines without any intervening absorber (hereafter non-absorber sightlines). The background quasars corresponding to the absorber and non-absorber sightlines lie within the redshift range of $0.421 \leq z_{\text{qso}} \leq 2.298$ and $0.388 \leq z_{\text{qso}} \leq 2.298$, respectively.

\begin{figure}
     \centering
     \includegraphics[width = 0.65\textwidth]{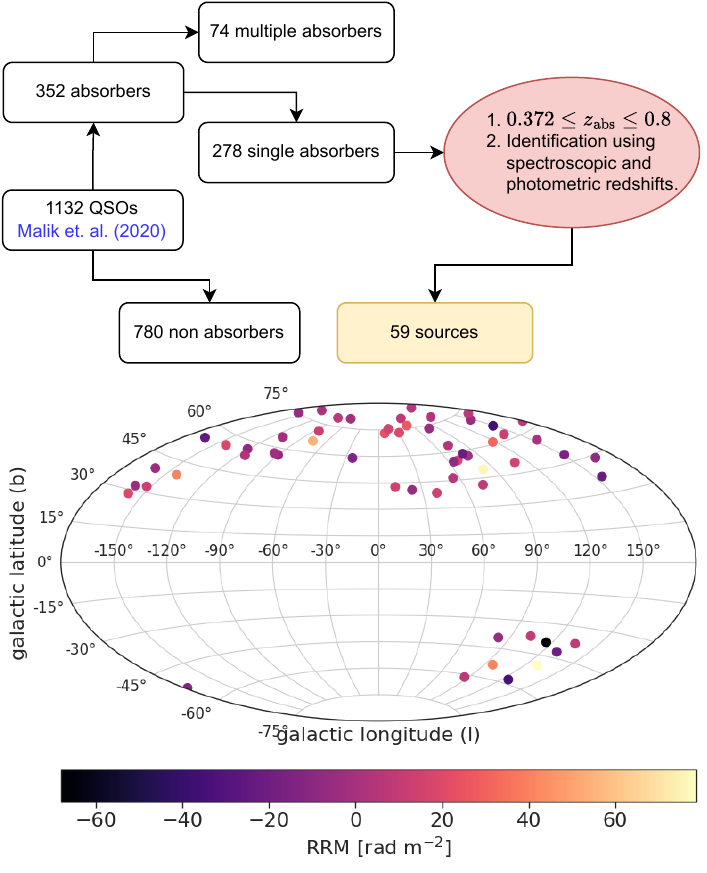}
     \caption{\textit{Top Panel:} Illustration of the sample generation procedure for extracting the final sample for 59 sightlines. \textit{Bottom Panel:} Location of the 59 galaxy-quasar pairs in the galactic coordinates, color-coded with the RRM values.}
     \label{fig:fig1}
\end{figure}
 
With the aim to infer the radial profile of the magnetic field in these intervening galaxies/absorbers, we take up the 352 sightlines having single/multiple absorbers in their foreground. As mentioned above, the absorbers can be in the redshift range of $0.372 \leq z_{\text{abs}} \leq 2.2$. We now lay down the detailed filtering procedure to extract our final sample of 59 sightlines having only one absorber (see top panel of Figure~\ref{fig:fig1}). The procedure is as follows: 
  
\begin{enumerate}
    \item To reduce the degeneracy among the contribution in excess RM from multiple absorbers, we selected 278 sightlines having a single absorber out of the initial sub-sample of 352 sources containing absorbers in the foreground.
    
    \item To ensure both the quasar and absorber galaxy are resolvable in the photometric data, we restricted our analysis to intervening galaxies with redshift $z_{\text{abs}} \leq 0.8$, narrowing down our selection to 112 sources within this redshift range. The upper limit is also constrained by the telescope's limited magnitude and spatial photometric resolution. 

     \item We retrieved the photometric images of these 112 sources from the SDSS DR16 Skyserver\footnote{\href{https://skyserver.sdss.org/dr16/en/home.aspx}{https://skyserver.sdss.org/dr16/en/home.aspx}} and searched for nearby galaxies with the photometric redshift consistent with the spectroscopic redshift to within a 3$\sigma_{\text{z-photo}}$ range. Here, $\sigma_{\text{z-photo}}$ represents the uncertainty in the photometric redshift of the galaxy. Using this technique, we successfully identified the associated galaxies for 59 out of the 112 sources.
\end{enumerate}

The bottom panel of Figure \ref{fig:fig1} shows the distribution of the above-generated sample in galactic coordinates. The sample consists of sources with galactic latitude, $|b| > 28^\circ$, which in addition to the GRM removal, has minimal effects of galactic foreground as discussed in \cite{Farnes_2014}. The 59 sources are associated with background quasars that lie within a redshift range of $0.452 \leq z_{\text{qso}} \leq 2.275$. More information regarding these 59 sources is listed in Table \ref{table:dataset_6_galaxies}. It is important to note that our selection process relied on a strict visual inspection, thereby ensuring the absence of any false detection of quasar-galaxy pairs.

\begin{table}
    \resizebox{\textwidth}{!}{%
    \centering
    \begin{tabular}{ccccccccccc}
        \hline
        \textbf{QSO} & \textbf{GAL} & \textbf{RM} & \textbf{$\delta\text{RM}$} & \textbf{RRM} & \textbf{$\delta\text{RRM}$} & \textbf{$\text{z}_\text{qso}$} & \textbf{$\text{z}_\text{abs}$} & \textbf{$\text{z}_\text{gal (photo)}$} & \textbf{$\sigma_{z-\text{photo}}$} & \textbf{D} \\ 
        \hline
        Jhhmmss.xx$\pm$ddmmss.xx & Jhhmmss.xx$\pm$ddmmss.xx & rad m$^{-2}$ & rad m$^{-2}$ & rad m$^{-2}$ & rad m$^{-2}$ & spec & spec & photo & photo & kpc \\ \hline
        J154459.43+040746.37 & J154459.45+040750.82 & 22.1 & 4.4 & 14.4 & 4.6 & 2.1881 & 0.42 & 0.339  & 0.0959 & 25.43 \\ 
        J140918.82+645521.49 & J140918.26+645525.09 & 45.2 & 13.7 & 16.0 & 13.8 & 1.0312 & 0.431 & 0.362  & 0.0486 & 29.21 \\
        J234831.77+062459.69 & J234831.99+062504.33 & 38.8 & 4.8 & 40.3 & 5.0 & 1.5396 & 0.38 & 0.413  & 0.0863 & 30.66 \\ 
        J102444.80+191220.42 & J102445.07+191223.73 & 25.6 & 5.5 & 18.4 & 5.7 & 0.8275 & 0.528 & 0.554  & 0.0349 & 32.59 \\ 
        J135726.48+001542.42 & J135726.17+001539.73 & -13.3 & 10.6 & -14.6 & 10.7 & 0.6614 & 0.477 &  0.42  & 0.1112 & 32.95 \\
        J083052.09+241059.82 & J083052.51+241101.94 & 12.0 & 0.9 & -5.2 & 1.7 & 0.9414 & 0.525 & 0.419  & 0.0495 & 39.6 \\
        \vdots & \vdots & \vdots & \vdots & \vdots & \vdots & \vdots & \vdots & \vdots  & \vdots & \vdots \\
        \\
        \hline
    \end{tabular}%
    }
    \caption{The table shows the required parameters of the quasar-galaxy pairs in our sample. The last column represents the impact parameter, `D', between the LoS of the quasar from the center of the intervening galaxies, which is confirmed by photometric and spectroscopic observations. The complete table containing all 59 galaxy-quasar pairs is available in machine-readable form. 
    }
    \label{table:dataset_6_galaxies}
\end{table}

\section{Analysis and Results}\label{sec:analysis_results}

\subsection{Estimation of Impact Parameters}\label{subsec:estm_D}

\begin{figure}
\centering
\includegraphics[width = \textwidth]{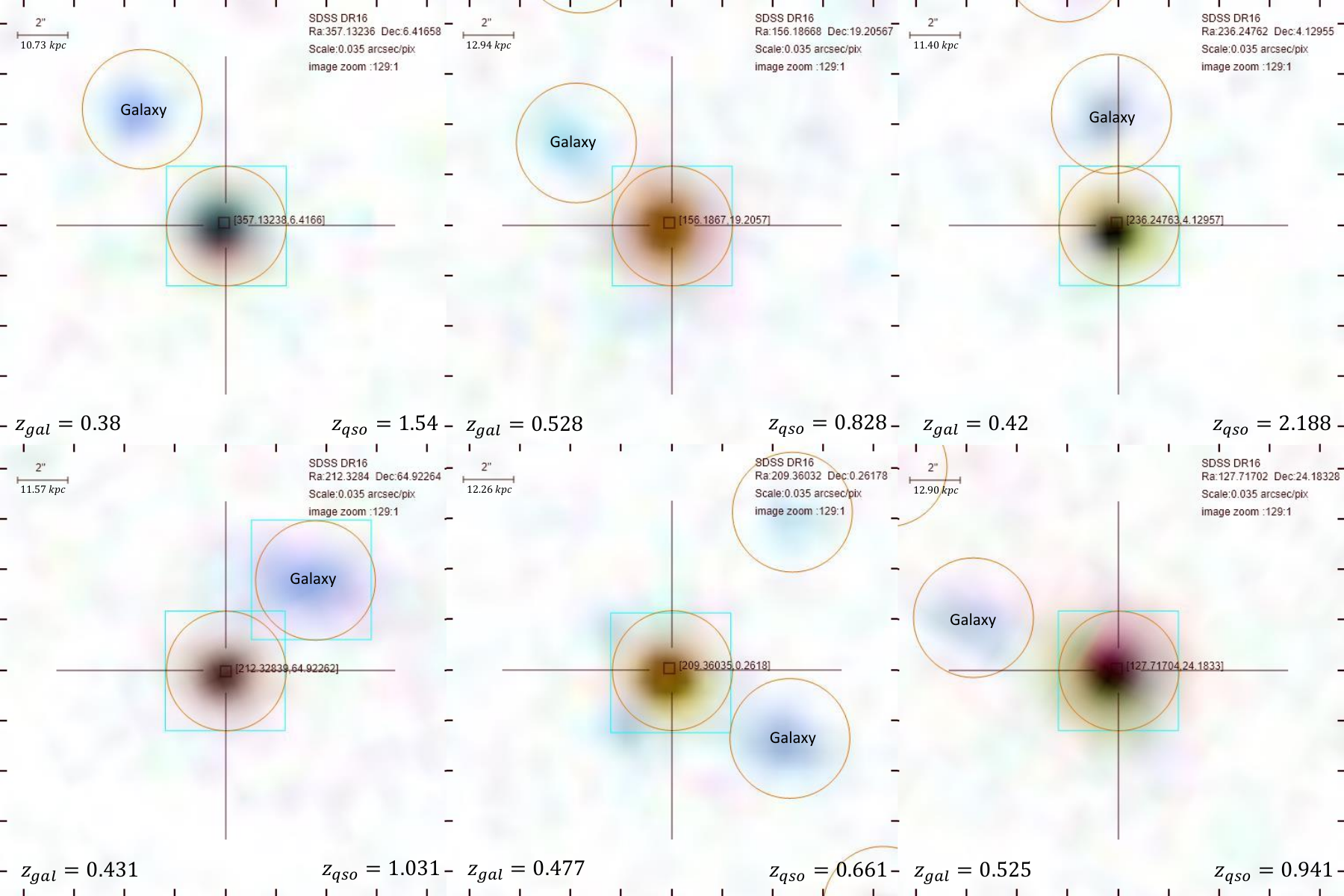}
\caption[]{The figure shows a sample of 6 sightlines from our sample of 59 sources as extracted from the SDSS Skyserver. These sub-panels of size $18''\times18''$ represent the region of the sky with a quasar in the center. We have taken the closest photometrically identified galaxy in the region (representing our intervening galaxy). We need to note that the sub-sample shown here is quasar-galaxy pairs with low-impact parameters ($\leq$ 40 kpc). Consequently, all the physical quantities, such as impact parameters, are calculated with respect to the intervening galaxy redshift. The quasars are depicted using various colours, which can primarily be attributed to their luminosity, given that they exist at different redshifts. The transverse distance mentioned at the top-left corner of each subpanel is evaluated at the redshift of the galaxy.}
\label{fig:6_sources_img}
\end{figure}

As discussed in Section \ref{sec:intro}, the magnetic field profiles of many nearby galaxies in the local group of the Milky Way have been much studied in the literature (for a review see \citep{Beck_2015}). However, the detailed radial profile of the magnetic field for the galaxies at high redshift is not yet well developed. To develop a radial profile of the galactic magnetic field, including the CGM up to $160$ kpc, we have utilized our controlled sample of 59 quasar-galaxy pairs. These galaxies have been identified in spectroscopic absorption spectra of the background quasar through the Mg \rom{2} doublet absorption lines and confirmed by a visual check from the photometric observations. As it is clear from Figure~\ref{fig:6_sources_img},  we can determine the angular separation of the quasar sightline from the center of the galaxy. It is given by

\begin{align}
\theta = \sqrt{\left((\Delta\delta)^2 + (\Delta\alpha)^2 \cos^2(\delta_{QSO})\right)}    
\end{align}

where $\Delta\alpha$ and $\Delta\delta$ are obtained from the RA and DEC of the quasar LoS and the center of its foreground galaxy. Subsequently, the impact parameter $D = \theta d_{A}(z_{gal})$, wherein $d_A(z_{gal})$ denotes the angular diameter distance at the redshift of the absorber galaxy~\citep{nielsen_magiicat_2013}. To compute the impact parameter corresponding to the observed angular separation for our quasar-galaxy system at the redshift of the foreground galaxy, we employ the \texttt{LambdaCDM} cosmological model available in the {\emph{Astropy}\footnote{\href{https://www.astropy.org/}{https://www.astropy.org/}}} library, following the framework presented by \cite{hogg_2000}. The key cosmological parameters utilized in this model are $H_0 = 67.74$ km s$^{-1}$ Mpc$^{-1}$, $\Omega_m = 0.31$, and $\Omega_\Lambda = 0.69$ \citep{Planck_2015}. Furthermore, there can be several sources of uncertainties in the impact parameters: I. Galaxy/Absorber's Redshift: The uncertainty in the absorber's redshift is very small compared to the bin width used in the analysis, and thus, it would not significantly affect the results. II. Spatial Uncertainties: The uncertainties from the astrometric/spatial resolution are not available for our dataset. However, if we approximate the spatial resolution of the telescope and translate it into the physical distance at the galaxy redshifts, it can lead to maximum uncertainties of less than 9.6 kpc in the impact parameters. This is within the 1$\sigma$ dispersion considered in our analysis. III. Cosmological Parameters: The choice of cosmological parameters will not introduce additional uncertainties but can shift the values of the impact parameters. Overall, the impact of these uncertainties on our analysis is minimal and within acceptable limits.

\begin{figure}
     \centering
     \includegraphics[width = 0.65\textwidth]{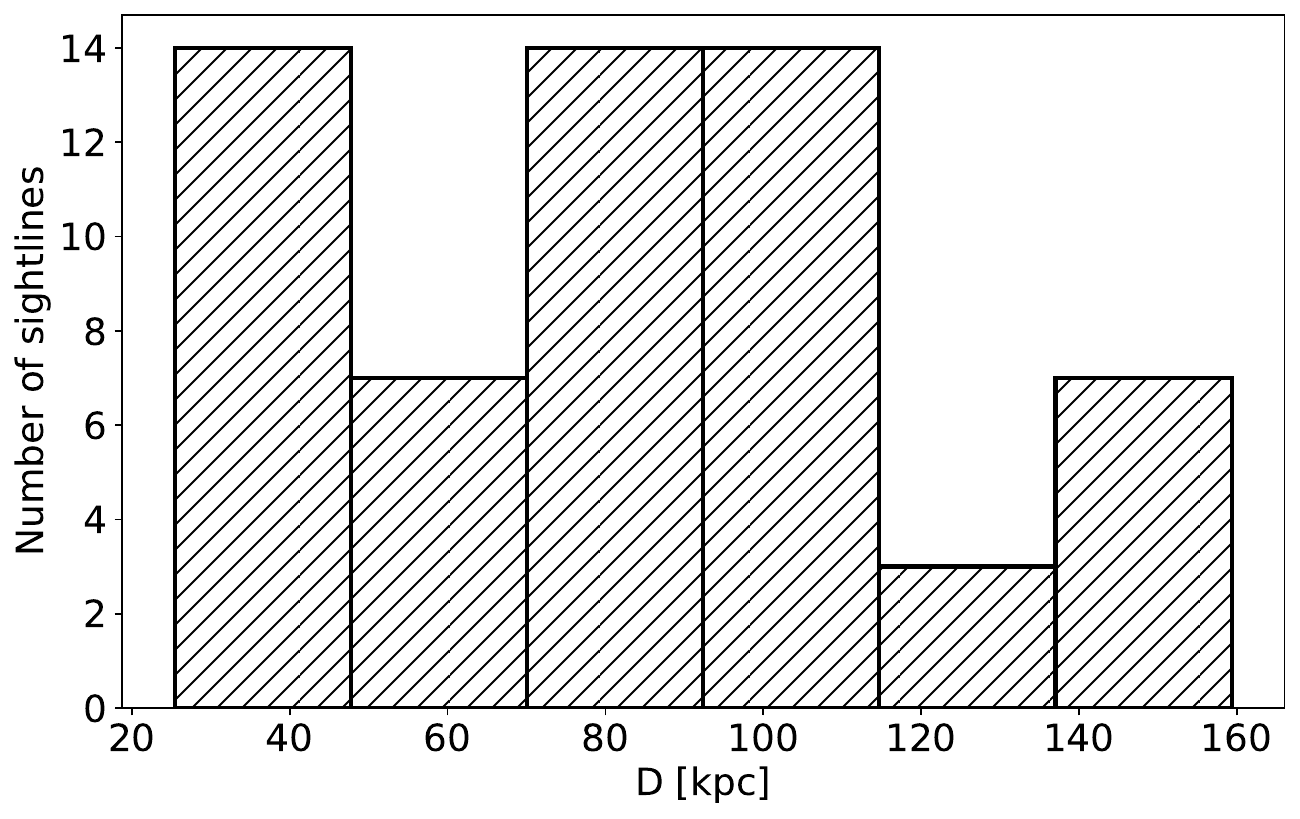}
     \caption{The figure shows the histogram (with a bin size of 22.3 kpc) of the 59 sources at different impact parameters.}
     \label{fig:D_hist_59}
 \end{figure}

Our sample encompasses a diverse range of impact parameters, from 25 kpc to 160 kpc, with a mean impact parameter of $84 \text{ kpc}$ (see Figure~\ref{fig:D_hist_59}). This range of the impact parameter corresponds to the CGM of typical Milky Way-type galaxies. 

\begin{figure}
    \centering
    \includegraphics[width = 0.65\textwidth]{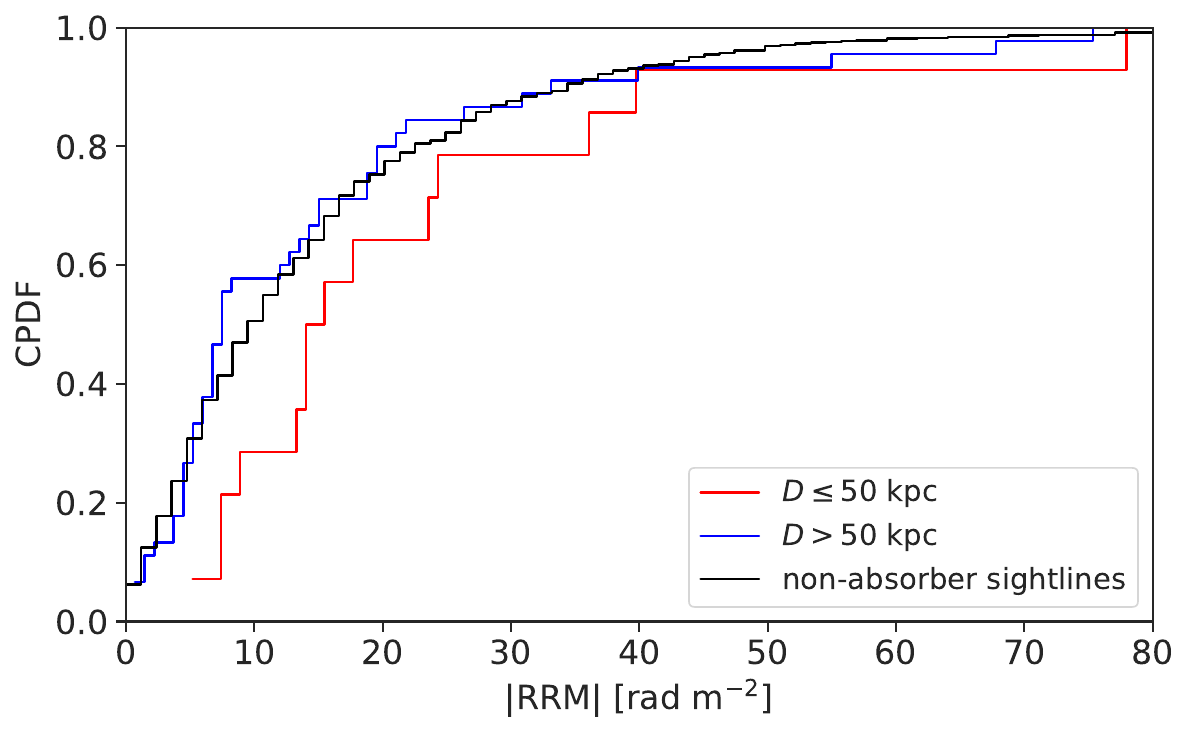}
    \caption{The figure shows the cumulative probability density function of the $|\text{RRM}|$ for two sub-samples binned at 50 kpc along with that of non-absorber sightlines. The CPDF of sightlines with $\text{D} \leq 50$ kpc is shown in red, whereas the CPDF for sightlines having $\text{D} > 50$ kpc and non-absorber sightlines are shown in blue and black, respectively. KS test for the three cases: (i) $\text{D} \leq 50$ and $\text{D} > 50;$ (ii) $\text{D} \leq 50$ and non-absorber sightlines; and (iii) $\text{D} > 50$ and non-absorber sightlines, rejects the null hypothesis at a confidence level of 95\%, 90\%, and 78\%, respectively.}
    \label{fig:cpdf_50_control}
\end{figure}

\begin{figure}
    \centering
    \includegraphics[width = 0.65\textwidth]{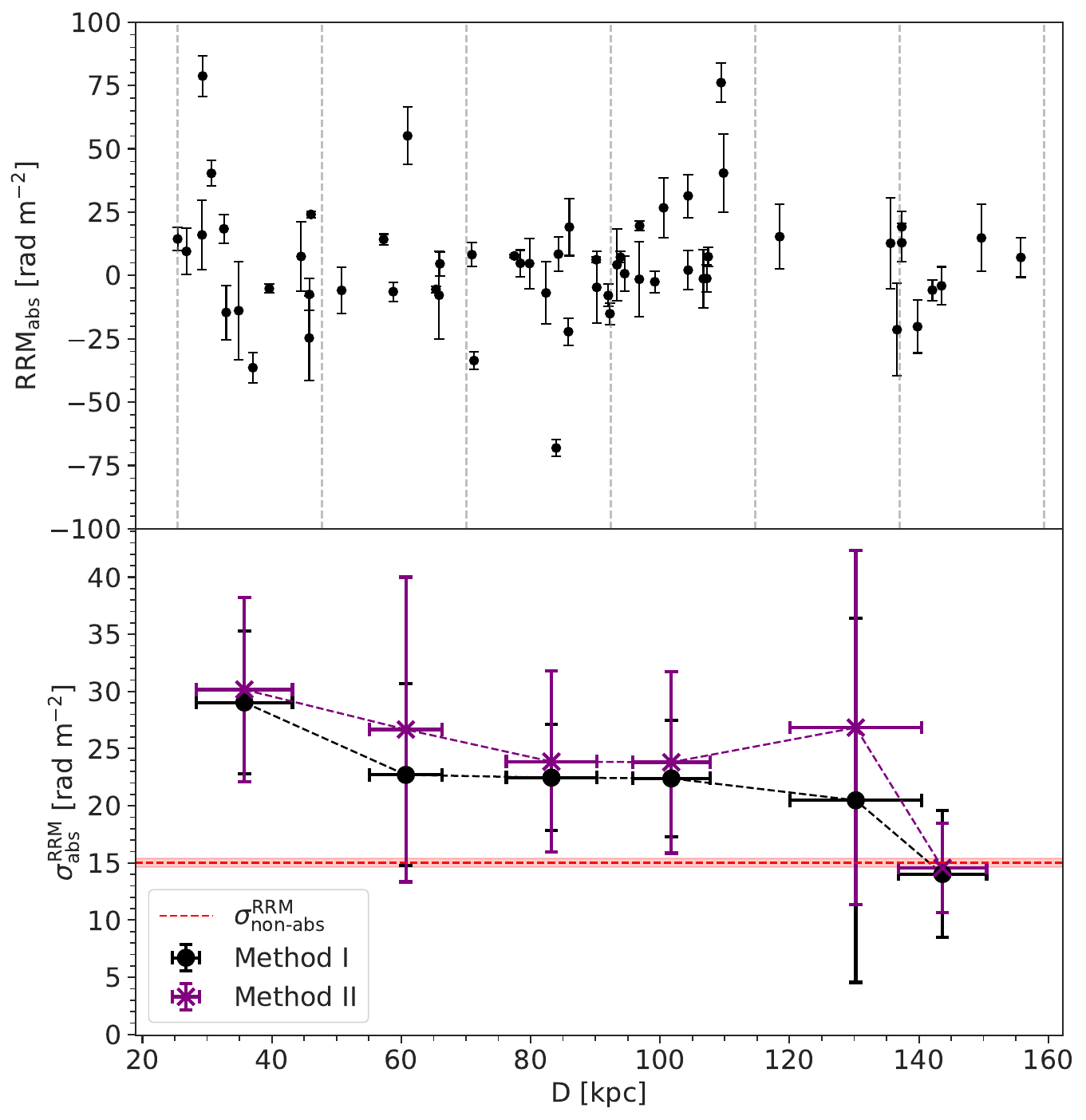}
    \caption{\textit{Top Panel:} The figure shows the distribution of the RRM of the sightlines with absorbers with the impact parameter (D) for our sample of 59 sightlines. \textit{Bottom Panel:} The plot shows the variation of ${\sigma_{\text{abs}}^{\text{RRM}}}$ calculated using Method I (black dots) and Method II (purple crosses), with D. The error bars on the dispersion are contributed by both the sampling error as well as the individual error on the data points. The horizontal red line with bands represents the $\sigma_{\text{non-abs}}^{\text{RRM}}= 15.01 \pm 0.6$ $\text{ rad } \text{m}^{-2}$. Since this represents the background dispersion coming from the sightlines with no absorbers, the error bars should be interpreted to be clipped by this threshold.}
    \label{fig:scatter_rrm}
\end{figure}
\subsection{Estimation of the Excess RRM and the radial profile}\label{subsec:rrm_excess_vs_D}
\label{radial}

As emphasized in Section \ref{sec:intro}, magnetic fields in any intervening system along the LoS to the source quasar will lead to an RRM excess over and above the IGM and the host contribution. We need to extract this carefully. To account for any redshift evolution in the RRM of quasars, we consider only those non-absorber sightlines with redshifts within the range $0.452 \leq z_{\text{qso}} \leq 2.275$. This criterion ensures that the redshift distribution of the selected non-absorber sightlines is consistent with that of the quasars with absorbers in our dataset. Following this redshift matching criterion, we obtain a sub-sample comprising 753 non-absorber sightlines. Utilizing the RRM and the impact parameters of our sample of 59 sightlines, we divide the sample into two sub-samples with impact parameters $\text{D}\leq50$ kpc and $\text{D}>50$ kpc. The standard deviations of RRM corresponding to these two sub-samples are $29.04 \pm 6.27$ $\text{rad } \text{m}^{-2}$ and $22.19 \pm 2.71$ $\text{rad } \text{m}^{-2}$, respectively. We performed the two-sampled Kolmogorov-Smirnov (KS) test\footnote{\href{https://real-statistics.com/non-parametric-tests/goodness-of-fit-tests/two-sample-kolmogorov-smirnov-test/}{https://real-statistics.com/non-parametric-tests/goodness-of-fit-tests/two-sample-kolmogorov-smirnov-test/}} \citep[see ][]{JL_Hodges_1958_ks2samp} to check the null hypothesis that these sub-samples are drawn from the same underlying population. A KS test rejects the null hypothesis at a 95\% confidence level. We also find that the KS tests for (i) $\text{D} \leq 50 \text{ kpc}$ with non-absorber sightlines, and (ii) $\text{D} > 50 \text{ kpc}$ with non-absorber sightlines rejects the null hypothesis with a confidence level of 90\%, and 78\%, respectively. Also, it can be seen from the cumulative probability density function in Figure \ref{fig:cpdf_50_control} that the sightlines having $\text{D} \leq 50$ kpc have greater RRM than the sightlines with $\text{D} > 50$ kpc and the non-absorber sightlines. This indicates that the RRM sub-samples have different characteristics with impact parameters and therefore can be used to estimate the radial profile. 

In the top panel of Figure \ref{fig:scatter_rrm}, we have plotted the distribution of $\text{RRM}$ for sightlines with absorbers, with the impact parameter. The plot suggests that there is a qualitative decrease in $\text{RRM}$ with the `radial' distance characterized by the impact parameter (assuming all intervenors are similar and thus this trend will be one for a typical galaxy in the chosen redshift range). To quantify the trend, we proceed by binning the sources in the impact parameter and computing the dispersion of RRM in each bin, $\sigma_\text{abs}^\text{RRM}$. There can be various approaches to compute $\sigma_\text{abs}^\text{RRM}$ and the uncertainity on the dispersion, $\delta\sigma_\text{abs}^\text{RRM}$. We have employed two procedures in our analysis, which we shall refer to as Method I and Method II hereafter. 
\begin{itemize}
\item \textbf{Method I}: In this method, we calculate $\sigma_\text{abs}^\text{RRM}$ characterized by the standard deviation of the sample and given by,

\begin{equation}
    \sigma_{\text{abs}}^\text{RRM} = \sqrt{\frac{\sum_{i = 1}^{N_b}{(\text{RRM}_i - \langle\text{RRM}\rangle)^2}}{N_b-1}}
    \label{eqn:standard_deviation}
\end{equation}
where $\langle\text{RRM}\rangle$ and $N_b$ represent the average RRM and number of sources, respectively, in each bin. 

The uncertainty  $\delta\sigma_\text{abs}^\text{RRM}$ in  $\sigma_\text{abs}^\text{RRM}$ has two contributions $\delta_1$ and $\delta_2$, where $\delta_1$ is estimated using the method of propagation of errors (see Appendix~\ref{appendix:error_estm}) which gives,
\begin{equation}
    \delta_1 = \frac{\sqrt{
      \sum_{i=1}^{N_b}{({\text{RRM}_i}-{\langle{\text{RRM}}\rangle})^2 \times ({\delta\text{RRM}_i})^2}
   }}{(N_b-1)\times\sigma_\text{abs}^\text{RRM}}
   \label{eqn:delta_sig_rm_intrv}
\end{equation}

where $\delta\text{RRM}_i$ is the propagated measurement uncertainty in individual RRM values. In addition to that, the sample dispersion gives the second contribution $\delta_2$ \citep[see][]{taylor1997introduction},
\begin{equation}
    \delta_2 = \frac{\sigma_{\text{abs}}^\text{RRM}}{\sqrt{2(N_b - 1)}}  
    \label{eqn:err_sig_obs}
\end{equation}
Therefore, the total error in the $\sigma_\text{abs}^\text{RRM}$, described by Eq.~\ref{eqn:standard_deviation}, is given by,
\begin{equation}
    \delta\sigma_{\text{abs}}^{\text{RRM}} = \sqrt{\delta_1^2 +  \delta_2^2}
    \label{eqn:error_total}
\end{equation}

\item \textbf{Method II}: In this method, we employ the jackknife resampling technique \citep[see][]{EfroTibs93, young2014wanted} to estimate the $\sigma_\text{abs}^\text{RRM}$ and $\delta\sigma_\text{abs}^\text{RRM}$. 
\end{itemize}

The bottom panel of Figure \ref{fig:scatter_rrm} shows the variation of $\sigma_\text{abs}^\text{RRM}$ obtained using Method I (black dots) and Method II (purple crosses) with impact parameter. The horizontal error bars represent the $1\sigma$ standard deviation of the impact parameters in each bin. We observe a decreasing trend in the $\sigma^{\text{RRM}}_{\text{abs}}$ with the impact parameter for both the methods, this is obvious if we consider all the bins except the fifth, the latter shows inconsistencies and large error bars on $\sigma_\text{abs}^\text{RRM}$ as a result of having only three data points in that bin, which significantly affects the jackknife resampling. The red dotted line represents the dispersion threshold $\sigma_{\text{non-abs}}^{\text{RRM}}= 15.01 \pm 0.6$ $\text{ rad } \text{m}^{-2}$ of the 753 non-absorber sightlines. 

Using the dispersion of the two samples, we estimate the excess RRM due to absorbers $\sigma_{\text{ex}}^{\text{RRM}}$ \citep{joshi_dependence_2013} as
\begin{equation}
    \sigma_{\text{ex}}^{\text{RRM}} = \sqrt{{(\sigma_{\text{abs}}^{\text{RRM}})}^2 -  {(\sigma_{\text{non-abs}}^{\text{RRM}})}^2}
    \label{eqn:v2_sig_rm_excess}
\end{equation}
where the error on $\sigma_{\text{ex}}^{\text{RRM}}$ given by 
\begin{equation}
    \delta{\sigma_{\text{ex}}^{\text{RRM}}} = \frac{1}{\sigma_{\text{ex}}^{\text{RRM}}}\sqrt{{(\sigma_{\text{abs}}^{\text{RRM}})}^2 (\delta{\sigma_{\text{abs}}^{\text{RRM}})}^2 +  {(\sigma_{\text{non-abs}}^{\text{RRM}})}^2 (\delta{\sigma_{\text{non-abs}}^{\text{RRM}})}^2}
    \label{eqn:v2_error_sig_rm_intrv}
\end{equation}

\begin{figure}
\centering
\includegraphics[width = 0.65\textwidth]{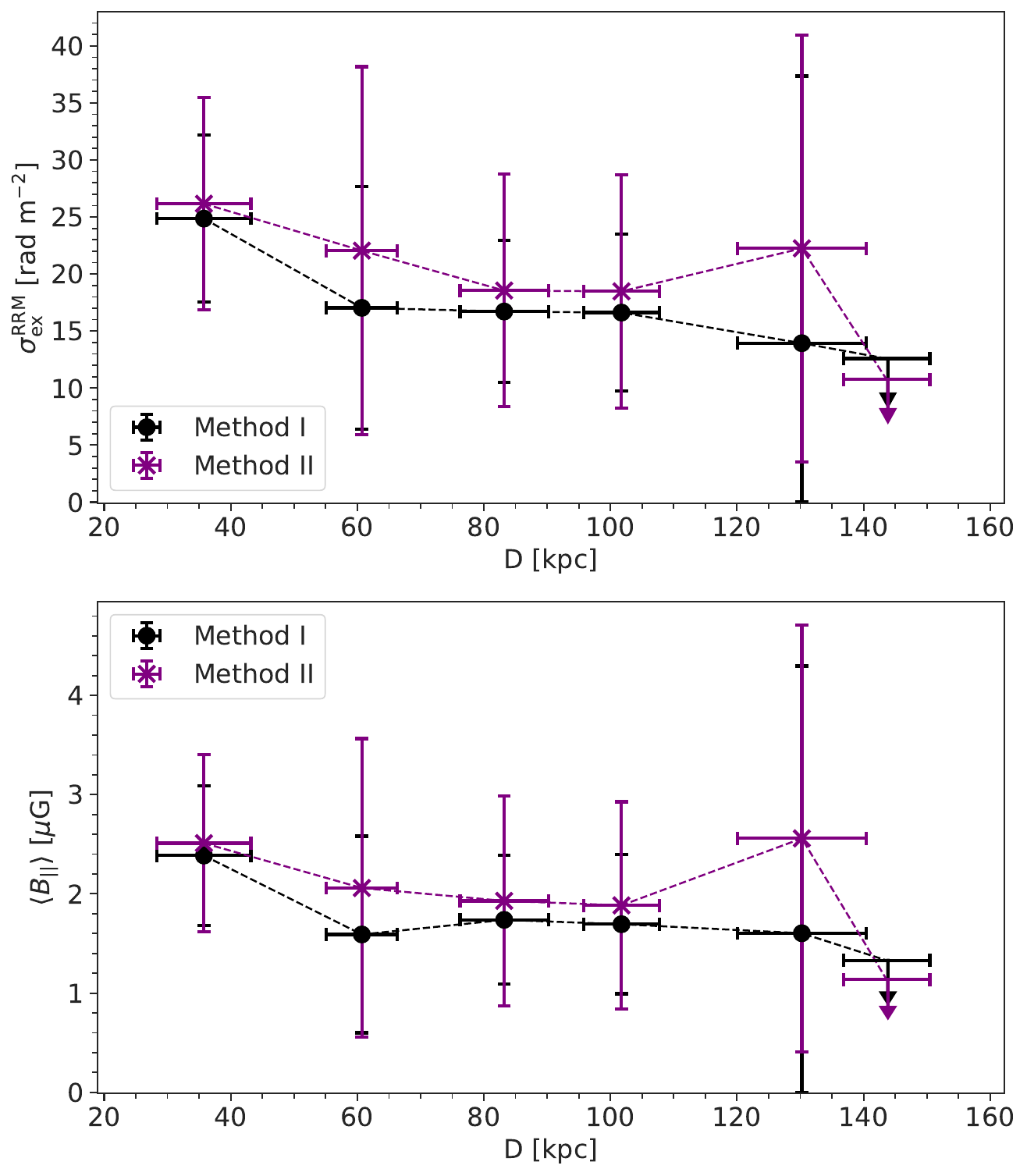}
\caption{\textit{Top Panel:} The figure shows the variation of the $\sigma_\text{ex}^\text{RRM}$ with the impact parameter. The black dots and purple crosses indicate the excess RRM obtained using $\sigma_\text{abs}^\text{RRM}$ from Method I and Method II, respectively. \textit{Bottom Panel:} The plot shows the derived strength of the LoS component of the magnetic field using $\sigma^{\text{RRM}}_\text{ex}$. We note that $\langle B_{||} \rangle$ versus D has Pearson Correlation Coefficient $\rho_p = -0.80$ and $-0.51$ for Method I and II, respectively. Note that $\langle B_{||} \rangle$ being field strength will always be positive definite; hence, we have asymmetric error bars for the last two points.}
\label{fig:sig_excess_B_field_59}
\end{figure}

The upper panel of Figure~\ref{fig:sig_excess_B_field_59} shows the variation of the excess RRM with impact parameter. For the last impact parameter bin (140-160 kpc), we place an upper bound on the $\sigma_{\text{ex}}^{\text{RRM}}$ using the upper limit of the $1\sigma$ error bar, that is, $\sigma^{\text{RRM}}_{\text{abs}} + \delta\sigma^{\text{RRM}}_{\text{abs}}$.
We observe that the $\sigma_{\text{ex}}^{\text{RRM}}$ traces a similar profile as of ${\sigma_{\text{abs}}^{\text{RRM}}}$ with the impact parameter along with the previously stated pathology in the penultimate bin. Also, we find $\sigma_{\text{ex}}^{\text{RRM}}$ for the two sub-samples with $\text{D} \leq 50$ kpc and $\text{D} > 50$ kpc to be $24.86 \pm 7.32$ rad m$^{-2}$ and $16.34 \pm 3.7$ rad m$^{-2}$, respectively. This decreasing radial trend in our analysis is also in line with the investigation by \cite{bernet_extent_2013} and intervening galaxy rotation measure as examined by~\cite{Lan_and_Prochaska_2020}. Recently, \cite{Heesen_CGM_magfield} and \cite{Bockmann_2023} also investigated the variation of rotation measure with the impact parameter to very large distances in the CGM. The authors found decreasing $|\text{RM}|$ profiles using larger bin sizes, whereas our analysis probes much finer physical length scales.

\subsection{Radial Profile of the Magnetic Field}\label{subsec:estm_B}

We obtain the radial profile of the galactic magnetic field using the relationship between RRM and the LoS component of the magnetic field strength. The column-averaged strength of the LoS magnetic field is given by~\citep[see Eq. 15 in][]{kronberg_global_2008}
\begin{equation}
\langle B_{\|} \rangle = 0.55 \mu \text{G}\left( \frac{1+z_{\text{med}}}{3.5} \right)^2 \left(\frac{\sigma_{\text{ex}}^{\text{RRM}}}{20 \ \text{rad} \ \text{m}^{-2}}\right) \left(\frac{N_e}{1.7 \times 10^{21} \text{cm}^{-2}}\right)^{-1}
\label{eqn:mag_field_from_sig_rm_intrv}
\end{equation}
where $z_{\text{med}}$ is the median redshift of the intervening galaxies lying in a given bin of the impact parameter, and $N_e$ denotes the electron column density. 

We assume similar properties among all the intervening galaxies within our sample of 59 sightlines, treating them as a single galactic system. For the electron column density, we adopted a value of $N_e \approx 9 \times 10^{19}$ cm$^{-2}$ as reported in \cite{Bernet2008, bernet_extent_2013} for a typical galaxy. Since the distances probes in our analysis from disk to CGM in the range of 25-160 kpc are significantly larger, considering the constant electron column density over these distances can be a fair approximation. Further, it is also supported by the studies carried out by \cite{constant_num_density_afruni_2017} and \cite{van_de_Voort_2018}.

Utilizing $\sigma_{\text{ex}}^{\text{RRM}}$, we obtain the radial profile of the galactic magnetic field $\langle B_{\|}\rangle$ as depicted in the bottom panel of Figure~\ref{fig:sig_excess_B_field_59}. We noted that the $\langle B_{||} \rangle$ is having anticorrelation with D, which can be quantified by the Pearson correlation coefficient $\rho_p = -0.80$ at a confidence limit of $\sim 95$\% (for Method I) and $\rho_p = -0.51$ at a confidence limit of 70\% (for Method II).   
Further, we have observed that the CGM exhibits a magnetic field with strengths exceeding $2 \mu$G near the galactic disc ($\sim 35$ kpc), and these strengths decrease to $\sim 1 \mu$G at greater distances from the galactic disc. This observation aligns well with the findings of \cite{Pakmor_CGM_2020MNRAS}, where the authors conducted a similar examination in simulation studies. Furthermore, a recent study by \cite{jung2023sampling}, employing the large cosmological simulation IllustrisTNG50 and a comprehensive galaxy catalog, has probed the magnetic field in high-velocity clouds within the CGM. This study reports substantial magnetic field strengths in these clouds with a radial decrease \citep[see Figure A1 in][]{jung2023sampling}. 
Our observational analysis hints the existence of such radial magnetic field profile within the CGM of the galaxies, which can be further explored in the future using more precise datasets. 

In comparing our observations with the aforementioned simulations, we must note that since the simulations are conducted in a 3D magnetic field model, a direct comparison alone may not suffice for drawing meaningful conclusions. 

Our analysis yields average LoS magnetic field strengths of $2.39 \pm 0.7 \ \mu$G and $1.67 \pm 0.38 \ \mu$G for sightlines having impact parameters $\text{D} \leq 50$ kpc and $\text{D} > 50$ kpc, respectively. This suggests a clear indication of varying magnetic fields from the disk to the CGM. 
The average magnetic field estimated from the excess RRM for the 59 sightlines is $\langle B_{\|}\rangle \sim 1.88 \pm 0.33 \ \mu$G at a median redshift of $z_{\text{med}} = 0.55$. Additionally, our estimated average magnetic field for the CGM is consistent with the constraints imposed by \cite{Lan_and_Prochaska_2020}.

We note that the obtained impact parameters primarily sample the CGM of the galaxy. Visual inspections and the SDSS classification revealed no complexity in the foreground galaxies, such as merging systems. Therefore, we have assumed that inclination and morphological effects are minimal in this first stride. We plan to include these effects in future investigations.

\subsection{Validation using simulated realizations}
\label{subsec:simulated_data}

To further investigate the effect of error in individual RRM values on the dispersion, we have checked into the simulated RRM observations. The procedure for generating synthetic observations is given below:

\begin{enumerate}

\item In real observations, we have 59 sightlines with given RRM and their respective error $\delta \text{RRM}$. We generated a sample of 1000 data points corresponding to each sightline assuming a Gaussian distribution with mean and standard deviation given by the RRM and $\delta \text{RRM}$ of sightline, respectively. 

\item As each of the 59 sightlines is associated with an impact parameter, we hence have 1000 realizations corresponding to each impact parameter from the above step. We randomly selected one simulated RRM from the respective set of 1000 values, for every original sightline, bin by bin in the impact parameter. For example, in the first impact parameter bin, we had originally 14 sources, which in turn, gave us 1000 simulated RRM values each. We selected an RRM value at random from each of these 1000 values to obtain 14 simulated values for the first bin. This procedure is repeated for all the bins. 

\item We computed the dispersion in the simulated RRM values, denoted by $\sigma^{\text{RRM}}_{\text{abs (sim)}}$. We repeated step (ii) to have 1000 realizations of $\sigma^{\text{RRM}}_{\text{abs (sim)}}$, and finally computed the average of these 1000 values.
\end{enumerate}

\begin{figure}
\centering
\includegraphics[width = 0.65\textwidth]{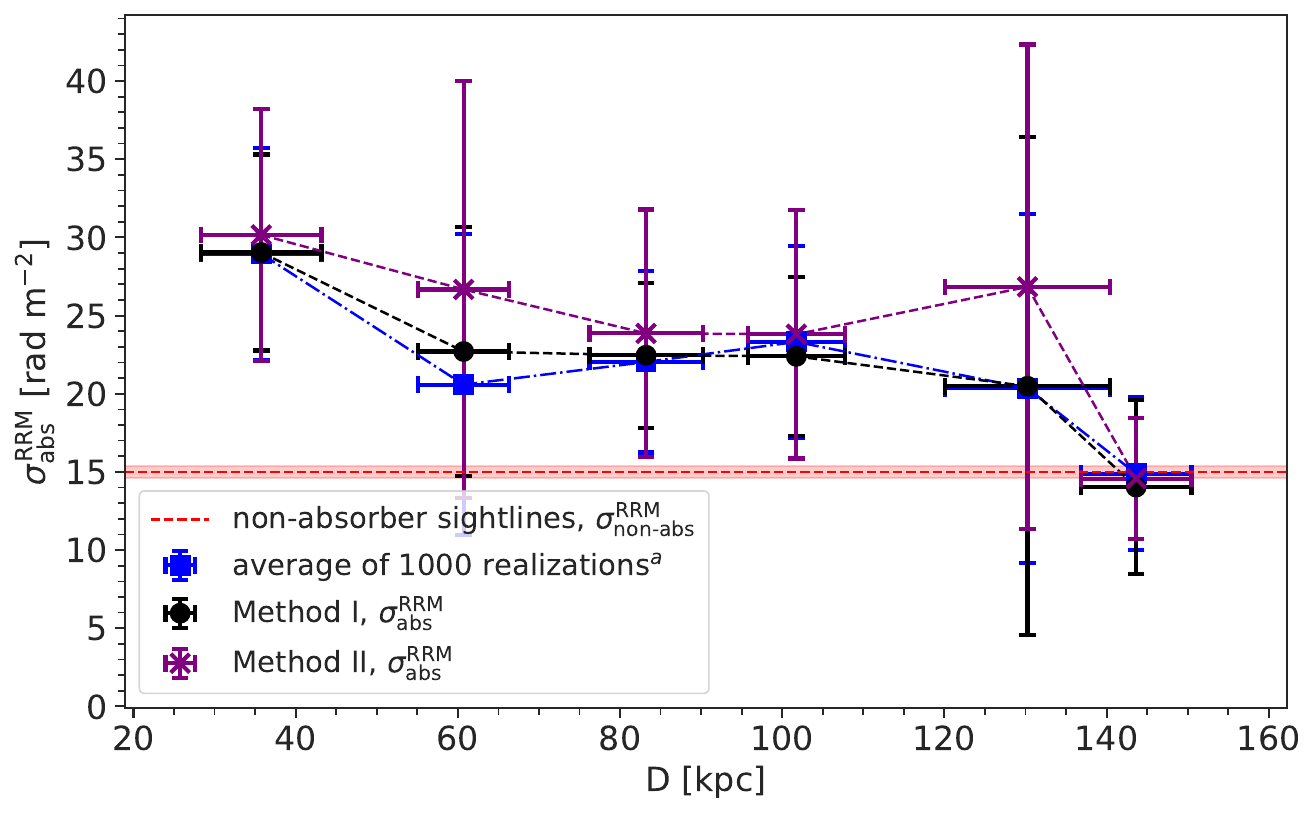}
\caption{The figure shows the variation of ${\sigma_{\text{abs}}^{\text{RRM}}}$ with the impact parameter D for three different cases (i) simulated datasets (blue squares), (ii) real observations Method I (black dots) and (iii) real observations Method II (purple crosses).  $^\text{a}$ We have estimated ${\sigma_{\text{abs}}^{\text{RRM}}}$ for 1000 simulated realizations and calculated the average over all the realizations. The details of the simulated dataset are given in Section~\ref{subsec:simulated_data}.}
\label{fig:RRM_std_realizations}
\end{figure}

We show the averaged $\sigma^{\text{RRM}}_{\text{abs (sim)}}$ for each bin (blue squares) with the impact parameter, in comparison with our real data, that is $\sigma^{\text{RRM}}_{\text{abs}}$ obtained from Method I (black dots) and Method II (purple crosses) in Figure~\ref{fig:RRM_std_realizations}. 
The red line with band represents the $\sigma^{\text{RRM}}_{\text{non-abs}}$ with its uncertainty $\delta\sigma^{\text{RRM}}_{\text{non-abs}}$. We find that the simulated radial trend corroborates the results of Section~\ref{subsec:rrm_excess_vs_D} (within the error bars). The error bars on the simulated data points ($\delta\sigma^{\text{RRM}}_{\text{abs(sim)}}$) represent the $1\sigma$ (standard deviation) range of the simulated $\sigma^{\text{RRM}}_{\text{abs(sim)}}$ for each bin.

\section{Discussion}
\label{sec:discussion}

Magnetic fields in the high redshift galactic systems have been probed using the polarized background quasar sightlines and their rotation measures. As mentioned before, RM is an integrated effect with contributions from the host environments, any intervening systems, the IGM, and our own galaxy, the Milky Way. We have inferred the radial profile of the magnetic field in the CGM of a typical high redshift galaxy using an ensemble of quasar-galaxy pairs. Our methodology is an advancement over some of the previous strides in this direction. Our results are in-line with the investigations in the earlier studies and also put forward observational evidence to results in a recent study \citep{jung2023sampling} coming from the cosmological simulation studies. Before presenting the final conclusion of our study, we shall now briefly discuss these aspects.

In many previous investigations, the excess in RM (or RRM) due to the Mg \rom{2} absorbers has been used to estimate the strength of the LoS component of magnetic fields in high redshift galaxies. For instance, \cite{Bernet2008} utilized their 71 sources with RM measurements at 6 cm and reported magnetic fields of the order of about 10$\mu$G. Additionally, \cite{kronberg_global_2008} performed an analysis with 268 sources by considering the exclusion of the galactic contribution to RM. The authors estimated an order of magnitude to be approximately $1\mu$G. This discrepancy in the order estimate could be due to the different sizes of the dataset and the directional dependence of GRM which has not been considered by \cite{Bernet2008}. \cite{Bernet_2010} using the RM data at 21 cm from the catalog compiled by \cite{taylor_rotation_2009} also did not find any correlation of Mg \rom{2} absorbers with RM. This concern was mitigated in the subsequent studies; for example, \cite{Farnes_2014} examined 599 sources based on the radio morphology of the background quasars with RM values at 21 cm. The authors found that the intervening Mg \rom{2} systems show an excess RM in flat spectrum sources and host magnetic fields of average strength of approximately $1.8 \pm 0.4$ $\mu$G. Later \cite{malik_role_2020} compiled a comparatively larger catalog of 1132 sources using 21 cm RM data and obtained a significant excess in RRM associated with the Mg \rom{2} absorbers. The authors also carried out an analysis based on the spectral index and found an excess RRM in flat-spectrum sources, agreeing with \cite{Farnes_2014}. This removed the above discrepancy among the 6 cm and 21 cm RM data. However, \cite{Bernet2012} discussed the above discrepancy using their model of depolarisation due to inhomogeneous Faraday screens. The authors carried out their analysis with 54 sources from \cite{Bernet2008} and studied the wavelength-dependent depolarisation effects with 6cm and 21cm polarization data. The analysis showed that the presence of intervening absorbers is associated with depolarization. Also, \cite{kim_2016} thoroughly examined a sample of 49 sources from \cite{Bernet2008} by the technique of RM synthesis. Their results also established a connection between the intervening absorbers and depolarization. \cite{malik_role_2020} found no significant difference between the fractional polarisation of the sources with and without absorbers suggesting that the major contribution to depolarization is due to the environment of the host (also taken up in a recent article by \cite{taylor2023mightee}). By utilizing the 1132 sources, \cite{malik_role_2020} found the average field strength by using excess RRM to be approximately $1.3 \pm 0.3$ $\mu$G. Our finding supports these previous investigations. We determined that the average strength of the LoS magnetic field of $1.88 \pm 0.33$ $\mu$G. Our work goes ahead by also considering the contribution to RRM from the host environment, and advancing a methodology to remove the same to obtain purely the contribution from the intervening galaxy in terms of $\sigma_{\text{ex}}^{\text{RRM}}$.  Also, we undertook special care in the identification of the galaxies. All the sources have been analyzed carefully with a visual check, and no automated search has been involved. 

The structure of the galactic magnetic fields has been well probed in the case of the Milky Way and the nearby galaxies. On a scale of qualitative analysis, our findings agree well with the previous investigations on the nearby galaxies \citep{Beck_2015}. Our work also hints at the existence of a decreasing profile as obtained in the cosmological simulations performed by \citep[see refs.][]{Pakmor_2017MNRAS, Pakmor_CGM_2020MNRAS, jung2023sampling}. In some earlier attempts to probe the radial profile of the magnetic fields in the high-$z$ galaxies, for instance, \cite{bernet_extent_2013} inferred a rundown of $|$RM$|$ with an impact parameter distribution up to about 50 kpc for 28 sightlines using RM observations at 6 cm. We have avoided mixing our sample (RM observations at 21 cm) with their sources at the 6 cm wavelength to maintain the homogeneity in the frequency of our sample. Further, their analysis emphasizes that the RM beyond 50 kpc is negligible. However, the coverage of impact parameters in our sample is comparatively much larger and well supported by the fact that CGM has a very high covering fraction up to 200 kpc~\citep[see refs.][]{Bordoloi_2018, Lan_and_Prochaska_2020}. Recently, \cite{Bockmann_2023} have studied the RM of 125 high redshift galaxies from the COSMOS and XMM-LSS fields with a median redshift of the collective sample of 0.42. The authors analyzed their sources up to 400 kpc and found that the galaxies in the COSMOS field with $\text{D} < 133$ kpc show an excess in the median $|\mathrm{RM}|$  of $2.5 \pm 1.1$ $\text{ rad } \text{m}^{-2}$ as compared to the galaxies with $\text{D} > 133$ kpc, and for XMM-LSS, the galaxies with $\text{D} < 266$ kpc show an excess of $5.0 \pm 2.3$ $\text{ rad } \text{m}^{-2}$. Thus, both the samples collectively result in a redshift corrected excess of $5.6 \pm 2.3$ $\text{ rad } \text{m}^{-2}$, using which they obtained a magnetic field strength of 0.48 $\mu$G. Also, \cite{Heesen_CGM_magfield} have probed magnetic fields in the CGM of nearby galaxies up to 1000 kpc based on a classification of sources on the basis of the inclination of the quasars' position with respect to the minor axis of the galaxies. Only those galaxies having quasars near the minor axis ($<45^{\circ}$) show a steep decrease in the $|$RRM$|$ up to 100 kpc. A similar analysis was carried out by \cite{Bordoloi_2011}, exploring the azimuthal profiles of Mg \rom{2} absorption. This indicates the presence of magnetized bipolar winds in the galaxies. Due to the low spatial resolution in SDSS at high redshifts, it is difficult to obtain any information about the morphology and orientation of our absorber galaxies. High-resolution observations are required to carry out an investigation on the bipolar winds in high redshift galaxies, which we will be using in follow-up work.
 
We do note that finer details in our analysis are also limited by the current dataset of RM observations. However, it is anticipated that future polarimetric surveys, such as the Square Kilometre Array \citep{SKA_galaxies8030053}, will help by supplying much more accurate RM data with a significantly higher number density of radio sources. In addition, the new measurements from spectroscopic and photometric surveys such as the SDSS and Dark Energy Spectroscopic Instrument (DESI\footnote{\href{https://www.desi.lbl.gov/}{https://www.desi.lbl.gov/}}) \citep{DESI_2019} are anticipated to open up a new avenue towards comprehending the galactic magnetic fields.

\section{Conclusion}\label{sec:conclusion}

In summary, our study focused on developing a probe to investigate the morphology of magnetic fields in high-redshift galaxies. We achieved this by utilizing the rotation measure of quasars whose sightlines intersect with the magnetized plasma of foreground galaxies. The key findings are as follows:

\begin{enumerate}
\item We have developed a sample of optically confirmed 59 quasars having a single Mg \rom{2} absorber associated with intervening foreground galaxies within a redshift range of $0.372 \leq z_{\text{abs}} \leq 0.8$. We have verified these galaxies using photometric as well as spectroscopic redshift with the 3$\sigma_{\text{z-photo}}$ limit. 
    
\item To calculate the RM contribution of the intervening galaxies, we began with RRM (RRM = RM - GRM), which has the Milky Way contributions already removed for all the sightlines, with and without the absorbers. We then found the excess in the dispersion of RRM, i.e., $\sigma_{\text{ex}}^{\text{RRM}}$ by subtracting the absorber and non-absorber dispersion in quadrature. 

\item
After dividing the sample of 59 sightlines into two sub-samples with $\text{D} \leq 50$ kpc and $\text{D} > 50$ kpc, we found, the excess RRM quantified by $\sigma_{\text{ex}}^{\text{RRM}}$ to be $24.86 \pm 7.32$ rad m$^{-2}$ and $16.34 \pm 3.7$ rad m$^{-2}$, respectively.

\item  We translated the excess RRM into magnetic field $\langle B_{\|}\rangle$ considering a typical electron column density, $N_e \approx 9 \times 10^{19}$ cm$^{-2}$. 
In these two sub-samples with $\text{D} \leq 50$ kpc and $\text{D} > 50$ kpc, $\langle B_{\|}\rangle$ is found to be $2.39 \pm 0.7 \ \mu$G and $1.67 \pm 0.38 \ \mu$G, respectively. It is noted that the magnetic field is showing anticorrelation with the impact parameter which further can be quantified by Pearson correlation coefficients -0.80 and -0.51 for Method I and Method II with confidence levels of 95\% and 70\%, respectively.

\item We have put forward a methodology to investigate the radial profile of the excess RRM and hence the magnetic field. Our focus is on better sampling (which we have done with the available NVSS and SDSS catalogs) through identification of foreground galaxies and obtaining a better RRM contribution due to the galaxies and analyzing its radial distribution. We see an indication of a radially decreasing magnetic field in a typical galaxy from disk to CGM (shown in Figure~\ref{fig:sig_excess_B_field_59}) 

\end{enumerate}

Our study provides a comprehensive technique to investigate the magnetic field in high-redshift galaxies. It can serve as a valuable precursor for ongoing and forthcoming RM surveys, including VLASS, LOFAR, and SKA.  

\appendix

\section{Derivation of the propagated error in the standard deviation}\label{appendix:error_estm}
To clarify the expression we have used for the error on the standard deviation (Eq. \ref{eqn:delta_sig_rm_intrv}), here's a derivation with the error propagation technique. To start, we consider the following general expression (for independent variables $x_i$'s),
\begin{equation}
\begin{aligned}
    \delta f(x_1, x_2, \cdots) = \sqrt{\left(\frac{\partial f}{\partial x_1}{\delta x_1}\right)^2 + \left(\frac{\partial f}{\partial x_2}{\delta x_2}\right)^2 + \cdots}
\end{aligned}\label{eq:appendix_err_prop_deriv_0}
\end{equation}
Now, since the standard deviation is
\begin{equation}
\begin{aligned}
    \sigma &= \sqrt{\frac{\sum_{i = 1}^N{(x_i - \Bar{x})^2}}{N-1}} = \sqrt{\frac{\sum_{i = 1}^N{(Nx_i - \sum_{j = 1}^N{x_j} )^2}}{N^2(N-1)}}\\
    \Rightarrow \frac{\partial \sigma^2}{\partial x_k} &= \frac{2}{N^2(N-1)}\sum_{i = 1}^N\left[(Nx_i - \sum_{j = 1}^N{x_j}){\left(N\frac{\partial x_i}{\partial x_k} - \sum_{j=1}^N\frac{\partial x_j}{\partial x_k}\right)} \right]\\
    &= \frac{2}{N^2(N-1)}\sum_{i = 1}^N\left[(Nx_i - \sum_{j = 1}^N{x_j}){(N\delta_{i,k} - 1 )} \right]\\
    &= \frac{2}{N(N-1)}\sum_{i = 1}^N\left[(x_i - \Bar{x}){(N\delta_{i,k} - 1 )} \right]\\
    &= \frac{2}{N(N-1)}\sum_{i = 1}^N\left[ Nx_i\delta_{i,k} - x_i - N\Bar{x}\delta_{i,k} + \Bar{x}\right]\\
    &= \frac{2}{N(N-1)}\left[ Nx_k - N\Bar{x} - N\Bar{x} + N\Bar{x}\right] = \frac{2(x_k-\Bar{x})}{N-1}\\
\end{aligned}\label{eq:appendix_err_prop_deriv_1}
\end{equation}
where, $\delta_{i,k}$ is the Kronecker delta. Using the Eq.~\ref{eq:appendix_err_prop_deriv_0} and Eq.~\ref{eq:appendix_err_prop_deriv_1}, we get
\begin{equation}
\begin{aligned}
\delta \sigma = \frac{\delta (\sigma^2)}{2\sigma} = \frac{\sqrt{\sum_{k = 1}^{N}{(x_k - \Bar{x})^2(\delta x_k)^2}}}{(N-1)\times\sigma}
\end{aligned}
\end{equation}

\acknowledgments

The research of S.S. is supported by the Core Research Grant CRG/2021/003053 from the Science and Engineering Research Board, India. Research of S.B. is supported by the Insitute Fellowship from the Indian Institute of Technology Delhi, India.

Funding for the Sloan Digital Sky Survey V has been provided by the Alfred P. Sloan Foundation, the Heising-Simons Foundation, the National Science Foundation, and the Participating Institutions. SDSS acknowledges support and resources from the Center for High-Performance Computing at the University of Utah. The SDSS web site is \href{www.sdss.org}{www.sdss.org}. SDSS is managed by the Astrophysical Research Consortium for the Participating Institutions of the SDSS Collaboration, including the Carnegie Institution for Science, Chilean National Time Allocation Committee (CNTAC) ratified researchers, the Gotham Participation Group, Harvard University, Heidelberg University, The Johns Hopkins University, L’Ecole polytechnique fédérale de Lausanne (EPFL), Leibniz-Institut für Astrophysik Potsdam (AIP), Max-Planck-Institut für Astronomie (MPIA Heidelberg), Max-Planck-Institut für Extraterrestrische Physik (MPE), Nanjing University, National Astronomical Observatories of China (NAOC), New Mexico State University, The Ohio State University, Pennsylvania State University, Smithsonian Astrophysical Observatory, Space Telescope Science Institute (STScI), the Stellar Astrophysics Participation Group, Universidad Nacional Autónoma de México, University of Arizona, University of Colorado Boulder, University of Illinois at Urbana-Champaign, University of Toronto, University of Utah, University of Virginia, and Yale University.

This work made use of Astropy:\footnote{\href{https://www.astropy.org/}{http://www.astropy.org}} a community-developed core Python package and an ecosystem of tools and resources for astronomy \citep{astropy:2013, astropy:2018, astropy:2022}.


\bibliographystyle{JHEP}
\bibliography{arxiv_jcap}

\end{document}